\documentclass[pre,aps,preprint,showpacs,preprintnumbers,amsmath,amssymb,secnumroman,eqsecnum]{revtex4}

\usepackage{graphicx}
\usepackage{dcolumn}
\usepackage{bm}

\newcommand{\beq}{\begin{equation}}
\newcommand{\eeq}{\end{equation}}
\newcommand{\beqa}{\begin{eqnarray}}
\newcommand{\eeqa}{\end{eqnarray}}
\newcommand{\vc}[1]{\mbox{\boldmath $#1$}}

\newcommand{\vol}[1]{{\bf #1}}

\newcommand{\du}[1]{{\bf\sf #1}}

\begin{document}


\title{Swimming of an assembly of rigid spheres at low Reynolds number}

\author{B. U. Felderhof}

 \email{ufelder@physik.rwth-aachen.de}
\affiliation{Institut f\"ur Theorie der Statistischen Physik \\ RWTH Aachen University\\
Templergraben 55\\52056 Aachen\\ Germany\\
}%



\date{\today}

\begin{abstract}
A matrix formulation is derived for the calculation of the swimming speed and the power required for swimming of an assembly of rigid spheres immersed in a viscous fluid of infinite extent. The spheres may have arbitrary radii and may interact with elastic forces. The analysis is based on the Stokes mobility matrix of the set of spheres, defined in low Reynolds number hydrodynamics. For small amplitude swimming optimization of the swimming speed at given power leads to an eigenvalue problem. The method allows straightforward calculation of the swimming performance of structures modeled as assemblies of interacting rigid spheres.
\end{abstract}

\maketitle
\section{\label{I}Introduction}

There is considerable recent interest in the construction of microswimmers capable of locomotion in a viscous fluid \cite{1}-\cite{3}. In the case of microorganisms the periodic changes of shape resulting in overall swimming motion are generated internally \cite {4}. In some robotic microswimmers the distortions of the body are generated from the outside \cite{5}-\cite{7}. The theory of swimming at low Reynolds number is nontrivial, and it is desirable to have simple models for which questions of principle can be elucidated and explicit calculations performed.

One such model is the three-sphere model first introduced by Najafi and Golestanian \cite{8}. In the model three spheres are aligned and motions along the line are considered. Hydrodynamic interactions were approximated by the Oseen tensor. Detailed analytic results were derived by Golestanian and Ajdari \cite{9}. The model and generalizations thereof were studied in computer simulation by Earl et al. \cite{10}. Vladimirov \cite{10A} studied the generalization to $N$ collinear spheres by means of a two-timing method. The effect of elastic direct interactions was studied in a model with the motion generated by actuating forces \cite{11}.

In the following we study swimming of more general sphere assemblies. It is shown that the swimming velocity and the rate of dissipation are conveniently calculated in a matrix formalism. As a starting point we use the exact mobility matrix, based on the equations of Stokes flow, for an arbitrary number $N$ of spheres immersed in infinite viscous fluid. Since the mobility matrix depends only on relative coordinates, the kinematics of motion can be prescribed in $3N-3$-dimensional relative coordinate space. It is shown that a cyclic path in relative space leads to a net translational shift of the assembly in ordinary space. Matrix expressions are derived for the mean swimming velocity, defined as the shift divided by the period of time, and for the mean rate of dissipation.

For small amplitude motion the mean swimming velocity and the mean rate of dissipation are given by bilinear expressions in terms of the displacements in relative space. These can be written as expectation values of two hermitian matrices for a complex $3N-3$-dimensional displacement vector characterizing the swimming stroke. The ratio of the two expressions provides a natural measure of the efficiency of the stroke. Optimization of the efficiency leads to an eigenvalue problem.

Numerical calculations have shown that the equations of Stokesian dynamics for a set of spheres lead to numerical instability of the solution, unless the spheres are subject to attractive direct interactions keeping them together \cite{11}. For simplicity one can assume harmonic elastic interactions between the spheres, bound harmonically to sites on a given equilibrium structure. The matrix formulation is extended to include the case where the forces on the spheres are a sum of harmonic interactions and internal or external actuating forces. The method allows straightforward analysis of the swimming performance of structures modeled as assemblies of freely rotating rigid spheres.

As an illustration we consider the three-sphere model of Najafi and Golestanian \cite{8}. The method works straightforwardly. In particular it can be used to optimize the swimming stroke for the model. The calculations can be performed for improved approximations to the hydrodynamic interactions, beyond the Oseen monopole approximation.

\section{\label{II}Displacement and swimming velocity}

We consider a set of $N$ rigid spheres of radii $a_1,...,a_N$ immersed in a viscous
incompressible fluid of shear viscosity $\eta$. The fluid is of infinite extent in all directions. At low Reynolds
number and on a slow time scale the flow velocity
$\vc{v}$ and the pressure $p$ satisfy the
Stokes equations \cite{12}
\begin{equation}
\label{2.1}\eta\nabla^2\vc{v}-\nabla p=0,\qquad\nabla\cdot\vc{v}=0.
\end{equation}
The flow velocity $\vc{v}$ is assumed to satisfy the no-slip boundary condition on the surface of the spheres.
The fluid is set in motion by time-dependent motions of the
spheres. At each time $t$ the velocity field $\vc{v}(\vc{r},t)$ tends to zero at infinity, and the pressure $p(\vc{r},t)$ tends to the constant ambient pressure $p_0$.
We shall study periodic relative motions which lead to swimming
motion of the collection of spheres.

We assume that the motion is caused by time-dependent periodic forces $\vc{F}_1(t),...,\vc{F}_N(t)$ which satisfy the condition that their sum vanishes at any time. The forces are transmitted by the spheres to the fluid. The spheres can rotate freely, so that they exert no torques on the fluid. Hence the rotational velocities $\vc{\Omega}_1(t),..., \vc{\Omega}_N(t)$ can be ignored. The translational velocities $\vc{U}_1,...,\vc{U}_N$ are linearly related to the forces,
\begin{equation}
\label{2.2}\vc{U}_j=\sum^N_{k=1}\vc{\mu}^{tt}_{jk}\cdot\vc{F}_k,\qquad j=1,...,N,
\end{equation}
 with translational mobility tensors $\vc{\mu}^{tt}_{jk}$. The tensors have many-body character and depend in principle on the positions of all particles  \cite{13}-\cite{15}. By translational invariance only relative distance vectors $\{\vc{R}_i-\vc{R}_j\}$ occur in the functional dependence. We abbreviate Eq. (2.2) as
 \begin{equation}
\label{2.3}\du{U}=\vc{\mu}\cdot\du{F},
\end{equation}
with a symmetric $3N\times 3N$ mobility matrix $\vc{\mu}$. Conversely
 \begin{equation}
\label{2.4}\du{F}=\vc{\zeta}\cdot\du{U},
\end{equation}
with friction matrix $\vc{\zeta}$. The friction matrix is the inverse of the mobility matrix, $\vc{\zeta}=\vc{\mu}^{-1}$,
and is also symmetric. Henceforth we omit the superscripts $tt$ for brevity.

The positions of the centers change as a function of time. The equations of motion of Stokesian dynamics read
  \begin{equation}
\label{2.5}\frac{d\vc{R}_j}{dt}=\vc{U}_j(\vc{R}_1,...,\vc{R}_N,t),\qquad j=1,...,N.
\end{equation}
The explicit time-dependence on the right originates in the time-dependence of the forces $\du{F}(t)$. We assume that the forces are periodic in time with period $T$, so that $\du{F}(t+T)=\du{F}(t)$. As mentioned, we impose the condition that at no time there is a net force acting on the set of spheres, so that
  \begin{equation}
\label{2.6}\sum^{N}_{j=1}\vc{F}_j(t)=0.
\end{equation}
We shall show that the solution of the set of nonlinear differential equations (2.5) takes the form
  \begin{equation}
\label{2.7}\vc{R}_j(t)=\vc{S}_j(t)+\vc{\xi}_j(t)=\vc{S}_{j0}+\overline{\vc{U}}_{sw}t+\vc{\xi}_j(t),\qquad j=1,...,N,
\end{equation}
where the positions $\{\vc{S}_j(t)\}$ describe the rigid body motion of a configuration $\du{S}_0=(\vc{S}_{10},...,\vc{S}_{N0})$ with mean swimming velocity $\overline{\vc{U}}_{sw}$, and the displacements $\vc{\xi}_j(t)$ are periodic in time, $\vc{\xi}_j(t+T)=\vc{\xi}_j(t)$. Hence $\overline{\vc{U}}_{sw}T$ is the net shift of a configuration in period $T$.

The condition Eq. (2.6) allows us to eliminate one of the forces, for example $\vc{F}_N$. Correspondingly the velocity $\vc{U}_N$ can also be eliminated. Since the mobility matrix depends only on relative coordinates, we can then reduce the spatial dimension of the algebraic problem. Eliminating $\vc{F}_N$ we define reduced velocities
  \begin{equation}
\label{2.8}\hat{\vc{U}}_j=\sum^{N-1}_{k=1}\hat{\vc{\mu}}_{jk}\cdot
\vc{F}_k,
\end{equation}
with reduced mobility tensors
  \begin{equation}
\label{2.9}\hat{\vc{\mu}}_{jk}=\vc{\mu}_{jk}-\vc{\mu}_{jN},\qquad j=1,...,N,\qquad k=1,...,N-1.
\end{equation}
Correspondingly we denote $\hat{\du{U}}=(\hat{\vc{U}}_1,...,\hat{\vc{U}}_{N-1})$ and $\hat{\du{F}}=(\vc{F}_1,...,\vc{F}_{N-1})$, and define the corresponding reduced $(3N-3)\times (3N-3)$ mobility matrix $\hat{\vc{\mu}}$ and friction matrix $\hat{\vc{\zeta}}$ by
  \begin{equation}
\label{2.10}\hat{\du{U}}=\hat{\vc{\mu}}\cdot\hat{\du{F}},\qquad\hat{\du{F}}=\hat{\vc{\zeta}}\cdot\hat{\du{U}},\qquad\hat{\vc{\zeta}}=\hat{\vc{\mu}}^{-1}.
\end{equation}
The matrices $\hat{\vc{\mu}}$ and $\hat{\vc{\zeta}}$ are not symmetric. The velocity of the $N$-th sphere is given by
  \begin{equation}
\label{2.11}\vc{U}_N=\sum^{N-1}_{j=1}\vc{\tau}_j\cdot\hat{\vc{U}_j},
\end{equation}
with dimensionless transfer tensors $\{\vc{\tau}_j\}$ given by
  \begin{equation}
\label{2.12}\vc{\tau}_j=\sum^{N-1}_{k=1}\hat{\vc{\mu}}_{Nk}\cdot\hat{\vc{\zeta}}_{kj},\qquad j=1,...,N-1.
\end{equation}

It is convenient to define relative coordinates $\{\vc{r}_j\}$ as
  \begin{eqnarray}
\label{2.13}\vc{r}_1&=&\vc{R}_2-\vc{R}_1,\qquad\vc{r}_2=\vc{R}_3-\vc{R}_2,\qquad ...,\nonumber\\
\vc{r}_{N-1}&=&\vc{R}_N-\vc{R}_{N-1}, \qquad j=1,...,N-1.
\end{eqnarray}
From the corresponding differentials we define displacements $\{\delta\vc{R}_j\}$ for $j=1,...,N-1$ as the solution of the equations
  \begin{eqnarray}
\label{2.14}d\vc{r}_1&=&\delta\vc{R}_2-\delta\vc{R}_1,\qquad d\vc{r}_2=\delta\vc{R}_3-\delta\vc{R}_2,\qquad ...,\nonumber\\
d\vc{r}_{N-1}&=&\sum^{N-1}_{j=1}\vc{\tau}_j\cdot\delta{\vc{R}_j}-\delta\vc{R}_{N-1}.
\end{eqnarray}
Via the tensors $\{\vc{\tau}_j\}$ the displacements $\{\delta\vc{R}_j\}$ depend on the relative coordinates $\{\vc{r}_j\}$, and therefore are uniquely defined at each accessible point of the $3N-3$-dimensional relative configuration space as linear combinations of the differentials $\{d\vc{r}_j\}$. In abbreviated notation Eq. (2.14) and its inverse read
  \begin{equation}
\label{2.15}d\du{r}=\du{S}\cdot\hat{\delta\du{R}},\qquad\hat{\delta\du{R}}=\du{Q}\cdot d\du{r},\qquad\du{Q}=\du{S}^{-1},
\end{equation}
with $\hat{\delta\du{R}}=(\delta\vc{R}_1,...,\delta\vc{R}_{N-1})$ and with $(3N-3)\times(3N-3)$-dimensional matrices $\du{S}$ and $\du{Q}$ which depend on the relative coordinates. We define in addition
 \begin{equation}
\label{2.16}\delta\vc{R}_N=\sum^{N-1}_{j=1}\vc{\tau}_j\cdot\delta{\vc{R}_j}.
\end{equation}

In the kinematic formulation of swimming the relative coordinates $\du{r}=\{\vc{r}_j\}$ and their time derivatives $d\du{r}/dt=\{d\vc{r}_j/dt\}$ are prescribed as periodic functions of time with frequency $\omega$. One cycle corresponds to a closed loop in $\du{r}$ space. It follows from Eqs. (2.14) and (2.16) that after a period the integral of the displacements over a cycle is independent of the label $j$, so that we may define the net displacement $\vc{\Delta}$ as
  \begin{equation}
\label{2.17}\vc{\Delta}=\oint\delta\vc{R}_j,\qquad j=1,...,N.
\end{equation}
The mean swimming velocity in Eq. (2.7) is given by
 \begin{equation}
\label{2.18}\overline{\vc{U}}_{sw}=\vc{\Delta}/T,
\end{equation}
with period $T=2\pi/\omega$. The displacement $\vc{\Delta}$ clearly is a geometrical property associated with radii $\{a_j\}$ and the Stokes equations Eq. (2.1). It may be evaluated for any given closed path in $\du{r}$-space and is independent of the nature of the prescribed time-dependence $\du{r}(t)$ on the path.

For given time-dependence $\du{r}(t)$ on a chosen closed path one may define the time-dependent displacements
 \begin{equation}
\label{2.19}\vc{\delta}_j(t)=\int^t_0\delta\vc{R}_j,\qquad j=1,...,N,
\end{equation}
and the corresponding mean value
 \begin{equation}
\label{2.20}\vc{\Delta}(t)=\sum^{N}_{j=1}p_j\vc{\delta}_j(t),\qquad\sum^{N}_{j=1}p_j=1,
\end{equation}
with positive weights $\{p_j\}$. In particular one may choose $p_j=m_j/\sum m_j$ corresponding to the center of mass, or $p_j=a_j/\sum a_j$ corresponding to the size distribution. The time-derivative $d\vc{\Delta}(t)/dt$ may be defined as the instantaneous swimming velocity $\vc{U}_{sw}(t)$. These quantities clearly depend on the choice of the weights $\{p_j\}$, but the net displacement $\vc{\Delta}=\vc{\Delta}(T)$ and the  mean swimming velocity $\overline{\vc{U}}_{sw}$ do not.

We have defined the displacement $\vc{\Delta}$ for the mobility matrix $\vc{\mu}(\du{r})$ corresponding to the set of radii and the no-slip boundary condition. For any closed path in $\du{r}$-space the displacement may also be calculated for an approximation to the mobility matrix, for example, as given by the Oseen-interaction. The same procedure may be followed  for any real $3N\times 3N$-dimensional matrix depending on $\du{r}$ for which the inverse of the corresponding matrix $\du{S}$ exists.

The spatial shift of the set of spheres during a period of the relative motion is related to the concept of geometric phase or holonomy \cite{11A}. Examples of holonomy are Foucault's pendulum, the four bar linkage studied by Yang and Krishnaprasad \cite{11B}, and Berry's phase in quantum mechanics \cite{11C}.

\section{\label{III}Dissipation and efficiency}

As the relative configuration of spheres varies as $\du{r}(t)$ runs through a cycle, the corresponding flow pattern in three-dimensional space also varies. The integral of the local rate of dissipation in the fluid at any time yields the total rate of dissipation $\mathcal{D}(t)$. The efficiency of swimming for the given cycle may be defined as the ratio of the net displacement $\vc{\Delta}$ and the mean rate of dissipation $\overline{\mathcal{D}}$, where the overline indicates the average over a period.

Alternatively the instantaneous rate of dissipation may be calculated from the forces and velocities as
 \begin{equation}
\label{3.1}\mathcal{D}(t)=\sum^{N}_{j=1}\vc{F}_j(t)\cdot\vc{U}_j(t).
\end{equation}
For the motion $\du{r}(t)$ we calculate velocities $\hat{\du{U}}(t)$ from Eq. (2.15) as
  \begin{equation}
\label{3.2}\hat{\du{U}}=\du{Q}\cdot \frac{d\du{r}}{dt}.
\end{equation}
The corresponding $N-1$ forces $\hat{\du{F}}(t)$ follow from Eq. (2.10). The velocity $\vc{U}_N(t)$ follows from Eq. (2.11) and $\vc{F}_N(t)$ is given by $-\sum\hat{\vc{F}}_j(t)$. The expression Eq. (3.1) may be cast in the form
 \begin{equation}
\label{3.3}\mathcal{D}(t)=\tilde{\frac{d\du{r}}{dt}}\cdot\tilde{\du{Q}}\tilde{\hat{\du{\zeta}}}\big[\du{I}-\du{U}_m\du{T}\big]\du{Q}\cdot\frac{d\du{r}}{dt},
\end{equation}
where the tilde denotes the transpose, $\du{I}$ is the $(3N-3)\times(3N-3)$ identity matrix, $\du{T}$ is a $(3N-3)\times(3N-3)$-dimensional matrix constructed from a $(N-1)\times(N-1)$-dimensional matrix with the tensors $\vc{\tau}_1,...,\vc{\tau}_{N-1}$ on the diagonal and zeros elsewhere, and $\du{U}_m$ is a $(3N-3)\times(3N-3)$-dimensional matrix given by
\begin{equation}
\label{3.4}\du{U}_m=\du{u}_x\du{u}_x+\du{u}_y\du{u}_y+\du{u}_z\du{u}_z,
\end{equation}
in Gibbs dyadic notation, where the symbol $\du{u}_x$ denotes a $3N-3$-dimensional vector with 1 on the $x$ positions, 0 on the $y,z$ positions, and cyclic. The antisymmetric part of the matrix drops out in Eq. (3.3) and we may use instead the symmetrized part
\begin{equation}
\label{3.5}\du{A}=\frac{1}{2\eta b}(\du{G}+\tilde{\du{G}}),\qquad\du{G}=\tilde{\du{Q}}\tilde{\hat{\vc{\zeta}}}\big[\du{I}-\du{U}_m\du{T}\big]\du{Q},
\end{equation}
where $b$ is the radius of the largest sphere. The matrix elements of $\du{A}$ are dimensionless. The rate of dissipation becomes
 \begin{equation}
\label{3.6}\mathcal{D}(t)=\eta b\tilde{\frac{d\du{r}}{dt}}\cdot\du{A}\cdot\frac{d\du{r}}{dt}.
\end{equation}
Since the rate of dissipation is positive definite, the matrix $\du{A}$ can be used to define a Riemannian metric in the accessible part of $\du{r}$-space with distance $ds$ given by
\begin{equation}
\label{3.7}ds^2=\tilde{d\du{r}}\cdot\du{A}\cdot d\du{r}.
\end{equation}

We consider in particular harmonically varying differentials of the form
 \begin{equation}
\label{3.8}d\du{r}(t)=\varepsilon[\hat{\vc{\xi}}_s\sin\omega t+\hat{\vc{\xi}}_c\cos\omega t],
\end{equation}
with infinitesimal factor $\varepsilon$ and constant vectors $\hat{\vc{\xi}}_s$ and $\hat{\vc{\xi}}_c$, which do not depend on $\du{r}$, and which characterize the stroke. Then at each accessible point $\du{r}$ the mean swimming velocity and the mean dissipation are at least of second order in $\varepsilon$,
\begin{equation}
\label{3.9}\overline{\vc{U}}_{sw}=\varepsilon^2\overline{\vc{U}}_{sw2}+O(\varepsilon^3),\qquad\overline{\mathcal{D}}=\varepsilon^2\overline{\mathcal{D}}_2+O(\varepsilon^3),
\end{equation}
with values which depend on $\hat{\vc{\xi}}_s,\;\hat{\vc{\xi}}_c$ and $\du{r}$. From Eqs. (2.18) and (3.8) we find
\begin{eqnarray}
\label{3.10}\overline{U}_{sw2\alpha}&=&\frac{1}{2}\omega\frac{\partial Q_{1\alpha,k\beta}}{\partial r_{l\gamma}}\big[\hat{\xi}_{cl\gamma}\hat{\xi}_{sk\beta}-\hat{\xi}_{sl\gamma}\hat{\xi}_{ck\beta}\big],\nonumber\\
\overline{\mathcal{D}}_2&=&\frac{1}{2}\eta b\omega^2\big[\hat{\vc{\xi}}_s\cdot\du{A}\cdot\hat{\vc{\xi}}_s+\hat{\vc{\xi}}_c\cdot\du{A}\cdot\hat{\vc{\xi}}_c\big].
\end{eqnarray}
The dimensionless efficiency for a particular stroke is defined as \cite{16},\cite{17}
\begin{equation}
\label{3.11}E_T=\eta\omega b^2\frac{|\overline{\vc{U}}_{sw2}|}{\overline{\mathcal{D}}_2}.
\end{equation}
This can be maximized with respect to the stroke $(\hat{\vc{\xi}}_s,\;\hat{\vc{\xi}}_c)$ at fixed values of the parameters. It was pointed out by Shapere and Wilczek \cite{16} that in the bilinear theory of swimming the above definition is preferable to Lighthill's efficiency \cite{19}, which is essentially the ratio of the square of velocity and the dissipation. In the bilinear theory the ratio of speed and power in Eq. (3.11) is independent of the amplitude of the stroke. At any amplitude, the stroke which maximizes the ratio of speed and power for chosen values of the parameters also maximizes Lighthill's efficiency, and minimizes the swimming drag coefficient of Avron et al. \cite{20}.

 \section{\label{IV}Optimizing efficiency}

 The quest for optimum efficiency of stroke at a point $\du{r}$ leads to an eigenvalue problem for the vectors $\hat{\vc{\xi}}_s$ and $\hat{\vc{\xi}}_c$. We abbreviate
\begin{equation}
\label{4.1}W^\alpha_{k\beta,l\gamma}=\frac{\partial Q_{1\alpha,k\beta}}{\partial r_{l\gamma}},
\end{equation}
 with a matrix $\du{W}^\alpha$ for $\alpha=(x,y,z)$. We have chosen sphere 1 for the definition, but this choice is arbitrary. Only the antisymmetric part of the matrix contributes to the swimming velocity in Eq. (3.10), so that we define the antisymmetric matrices $\du{Y}^\alpha$ by
  \begin{equation}
\label{4.2}\du{Y}^\alpha=\frac{1}{2}(\du{W}^\alpha-\tilde{\du{W}}^\alpha),\qquad \alpha=(x,y,z).
\end{equation}
The swimming velocity in Eq. (3.10) can then be expressed as
\begin{equation}
\label{4.3}\overline{U}_{sw2\alpha}=\omega\hat{\vc{\xi}}_s\cdot\du{Y}^\alpha\cdot\hat{\vc{\xi}}_c.
 \end{equation}
 The associated eigenvalue problem is expressed conveniently in complex notation. We introduce the complex dimensionless vector
 \begin{equation}
\label{4.4}\hat{\vc{\xi}}^c=\frac{1}{b}(\hat{\vc{\xi}}_c+i\hat{\vc{\xi}}_s),
 \end{equation}
 and the dimensionless hermitian matrices
  \begin{equation}
\label{4.5}\du{B}^\alpha=ib\du{Y}^\alpha,\qquad\alpha=(x,y,z).
 \end{equation}
We ask for the stroke with maximum swimming velocity in a class of strokes with equal rate of dissipation for fixed values of the geometric parameters, fixed frequency $\omega$, and fixed viscosity $\eta$. This leads to the eigenvalue problem
 \begin{equation}
\label{4.6}\du{B}^\alpha\hat{\vc{\xi}}^c=\lambda^\alpha\du{A}\hat{\vc{\xi}}^c.
 \end{equation}
 The eigenvalues $\{\lambda^\alpha\}$ are real. The real and imaginary parts of the eigenvectors yield the corresponding $\hat{\vc{\xi}}_c$ and $\hat{\vc{\xi}}_s$. The eigenvalues occur in pairs
 $\pm\lambda^\alpha$ with eigenvectors which are complex conjugate, corresponding to swimming in opposite directions. It is convenient to use the notation
 \begin{equation}
\label{4.7}\overline{U}_{sw2\alpha}=\frac{1}{2}\omega
b(\hat{\vc{\xi}}^c|\du{B}^{\alpha}|\hat{\vc{\xi}}^c),\qquad\overline{\mathcal{D}}_2=\frac{1}{2}\eta\omega^2b^3(\hat{\vc{\xi}}^c|\du{A}|\hat{\vc{\xi}}^c),
 \end{equation}
 with the scalar product
  \begin{equation}
\label{4.8}(\hat{\vc{\xi}}^c|\hat{\vc{\eta}}^c)=\sum^{N-1}_{j=1}\hat{\vc{\xi}}_j^{c*}\cdot\hat{\vc{\eta}}^c_j.
 \end{equation}
 The maximum efficiency for motion in direction $\alpha$ is given by the maximum eigenvalue as
   \begin{equation}
\label{4.9}E^\alpha_{Tmax}=\lambda^\alpha_{max}.
 \end{equation}
 The set $\{E^x_{Tmax},E^y_{Tmax},E^z_{Tmax}\}$ depends on the choice of Cartesian coordinate system. Further optimization may be possible by a rotation of axes. In particular cases a natural choice of axes will suggest itself.

 We recall that the matrices $\du{B}^\alpha$ and $\du{A}$ depend on the point in $\du{r}$-space under consideration. At a chosen point $\du{r}_0$ the maximum eigenvalue for optimized choice of axes yields the maximum swimming velocity for given power, and the corresponding eigenvector yields the nature of the corresponding stroke. The amplitude factor $\varepsilon^2$ in Eq. (3.9) implies that the amplitude must be small. It is natural to consider larger amplitudes by use of Eq. (3.8) and calculate the mean swimming velocity and rate of dissipation by use of Eqs. (2.18) and (3.6) for the corresponding cyclic path $\du{r}(t)$ centered about $\du{r}_0$.

\section{\label{V}Harmonic interactions and actuating forces}

In the formulation of the mobility matrix in Eq. (2.2) the nature of the forces $\{\vc{F}_j\}$ need not be specified. In an earlier calculation \cite{11} we have considered microswimmers with internal harmonic interactions, driven by actuating forces. The forces are of the form
  \begin{equation}
\label{5.1}\vc{F}_j=\vc{E}_j+\sum_{k=1}^N\vc{H}_{jk}\cdot(\vc{\xi}_j-\vc{\xi}_k),
 \end{equation}
 with displacements
  \begin{equation}
\label{5.2}\vc{\xi}_j(t)=\vc{R}_j(t)-\vc{S}_{j0}-\overline{\vc{U}}_{sw}t,
 \end{equation}
 where $\{\vc{S}_{j0}\}$ are equilibrium positions at time $t=0$.
The constant harmonic interaction tensors $\vc{H}_{jk}$ are symmetric and satisfy $\vc{H}_{jk}=\vc{H}_{kj}$.
The actuating forces $\{\vc{E}_j(t)\}$ are assumed to satisfy
 \begin{equation}
\label{5.3}\sum_{j=1}^N\vc{E}_j(t)=0.
 \end{equation}
They can be internal or generated from the outside. As seen from Eq. (2.13) the differences $\{\vc{\xi}_j-\vc{\xi}_k\}$ depend linearly on the relative coordinates $\du{r}$, so that for an equilibrium configuration with relative distances $\du{r}_0$ the first $N-1$ forces may be expressed as
 \begin{equation}
\label{5.4}\hat{\du{F}}=\hat{\du{E}}+\du{K}\cdot(\du{r}-\du{r}_0),
 \end{equation}
 with a real matrix $\du{K}$. It is convenient to rewrite the second order mean swimming velocity as a bilinear expression in terms of the displacements $\hat{\vc{\xi}}=\du{r}-\du{r}_0$ and forces $\hat{\du{F}}$. To simplify notation we agree that the subscripts $l,m,n,p,q$ comprise both particle label and Cartesian index. The subscripts $i,j,k$ are reserved for particle labels. It follows from Eqs. (2.15) and (2.17) that the mean second order swimming velocity can be expressed as
 \begin{equation}
\label{5.5}\overline{U}_{sw2\alpha}=\frac{1}{T}\int^T_0
dr_m\frac{\partial Q_{j\alpha,l}}{\partial
r_m}\bigg|_0\frac{dr_l}{dt}\;dt,\qquad j=1,...,N-1,
 \end{equation}
 where summation over repeated indices is implied. The first order equations of motion in relative coordinate space are given by
  \begin{equation}
\label{5.6}\frac{d\du{r}}{dt}=\frac{d\hat{\vc{\xi}}}{dt}=\du{L}\cdot\hat{\du{F}},\qquad\du{L}=\du{S}\hat{\vc{\mu}}\big|_0.
 \end{equation}
 Substituting into Eq. (5.5) we obtain for the swimming velocity
  \begin{equation}
\label{5.7}\overline{U}_{sw2\alpha}=\frac{1}{T}\int^T_0 (\hat{\vc{\xi}}(t)|\du{X}^{j\alpha}|\hat{\du{F}}(t))\;dt,\qquad j=1,...,N-1,
 \end{equation}
 where the matrix $\du{X}^{j\alpha}(\du{r}_0)$ is independent of time and has matrix elements
 \begin{equation}
\label{5.8}X^{j\alpha}_{mn}=\frac{\partial Q_{j\alpha,p}}{\partial
r_m}L_{pn}.
 \end{equation}
 The matrix is independent of the particle label $j$. Similarly, from Eq. (3.6) the second order mean rate of dissipation may be expressed as
   \begin{equation}
\label{5.9}\overline{\mathcal{D}}_2=\frac{\eta b}{T}\int^T_0 (\frac{d\hat{\vc{\xi}}}{dt}|\du{A}|\frac{d\hat{\vc{\xi}}}{dt})\;dt,
 \end{equation}
 with matrix $\du{A}$, given by Eq. (3.5), calculated at $\du{r}_0$.

 Substituting Eq. (5.4) into Eq. (5.6) and assuming harmonically varying actuating forces one obtains in complex notation
    \begin{equation}
\label{5.10}-i\omega b\hat{\vc{\xi}}^c_\omega=\du{L}\cdot(\hat{\du{E}}^c_\omega+b\du{K}\cdot\hat{\vc{\xi}}^c_\omega),
 \end{equation}
 with the solution
    \begin{equation}
\label{5.11}b\hat{\vc{\xi}}^c_\omega=\big[-i\omega\du{I}-\du{L}\du{K}\big]^{-1}\du{L}\cdot\hat{\du{E}}^c_\omega.
 \end{equation}
 This relation may be used to express the swimming velocity in Eq. (5.7) and the mean rate of dissipation in Eq. (5.9) in terms of the actuating forces $\hat{\du{E}}^c_\omega$. The resulting dynamic expression for the efficiency is maximized by the method of Sec. IV and calculation of the corresponding actuating force amplitudes $\hat{\du{E}}_s$ and $\hat{\du{E}}_c$, defined in analogy to Eq. (3.8), from Eq. (5.11).

 \section{\label{VI}Three-sphere swimmer}

 The simplest application of the theory is to a three-sphere swimmer with three spheres aligned on the $x$ axis, as studied by Golestanian and Ajdari \cite{9}. The arms mentioned by these authors do not appear in the theory, and the derived equations apply to three spheres in the absence of direct interactions. The authors define the instantaneous swimming velocity as the mean $V=(U_1+U_2+U_3)/3$. As seen above, this definition is somewhat arbitrary. One could also consider the center of mass or the center of sizes. As we have shown, for periodic motion the mean velocity of any sphere, averaged over a period, provides the mean swimming velocity.

 In the example the $y$ and $z$ coordinates can be ignored. There are only two relative coordinates $r_1=x_2-x_1$ and $r_2=x_3-x_2$, and the various matrices $\du{A},...,\du{Y}$ are two-dimensional. The elements of the $3\times 3$ mobility matrix are approximated by use of the Oseen interaction as \cite{12}
     \begin{equation}
\label{6.1}\mu^{tt}_{jk}=\frac{1}{6\pi\eta}\bigg[\frac{1}{a_j}\delta_{jk}+\frac{3}{2|x_j-x_k|}\bigg].
 \end{equation}
 In the bilinear theory we consider a point $\du{r}_0$ in $\du{r}$-space with coordinates $(d_1,d_2)$. It is not necessary to assume the ratios $a_j/d_k$ to be small, as done by Golestanian and Ajdari. The more complete expressions will provide a better approximation to the situation with exact hydrodynamic interactions. Similarly one could use the more complicated Rotne-Prager expressions \cite{18} instead of Eq. (6.1).

 As an example we consider the case of equal-sized spheres with $a_1=a_2=a_3=a$ and equal distances between centers $d_1=d_2=d$. The matrix $\du{B}^x$ in Eq. (4.5) takes the form
  \begin{equation}
\label{6.2}\du{B}^x=\left(\begin{array}{cc}0&iaY^x_{12}
\\-iaY^x_{12}&0
\end{array}\right),
\end{equation}
with element
  \begin{equation}
\label{6.3}Y^x_{12}=\frac{-a}{3d}\frac{56d^2-198da+189a^2}{(4d-3a)(4d-7a)^2}.
\end{equation}
We note that the corresponding swimming velocity given by Eq. (4.7) with $b=a$ when expanded to lowest order in the ratio $a/d$ agrees with Eq. (12) of ref. 9.
The matrix $\du{A}$ takes the form
 \begin{equation}
\label{6.4}\du{A}=\frac{8\pi d}{(4d-3a)(4d-7a)}\left(\begin{array}{cc}8d-12a&4d-9a
\\4d-9a&8d-12a
\end{array}\right),
\end{equation}
These expressions show a singularity at $d=7a/4$, which is less than the minimum distance $2a$. The matrix $\du{Q}$, defined in Eq. (2.15), is singular at $d=7a/4$. From Eq. (4.6) one finds the eigenvalues
\begin{equation}
\label{6.5}\lambda_\pm=\mp\frac{a}{8\sqrt{3}\pi d}\sqrt{(4d-3a)(4d-7a)}Y^x_{12},
\end{equation}
 as well as the corresponding eigenvectors $\vc{\xi}_\pm=(1,\xi_\pm)$ with
 \begin{equation}
\label{6.6}\xi_+=\frac{1}{8d-12a}\bigg[-4d+9a+i\sqrt{3(4d-3a)(4d-7a)}\bigg],\qquad\xi_-=\xi_+^*,
\end{equation}
normalized to $(\vc{\xi}_+|\vc{\xi}_+)=2$. The maximum efficiency, corresponding to $\lambda_+$ by Eq. (4.9), tends to zero monotonically as the ratio $d/a$ tends to infinity.

 It is of interest to compare the above analytical results for small amplitude motion with values obtained by numerical solution of the equations of motion Eq. (2.5) with hydrodynamic interactions given by Eq. (6.1) and prescribed oscillating forces. As we have noted earlier \cite{11}, the numerical solution is unstable unless one introduces harmonic elastic forces as in Eq. (5.4), keeping the particles together. With suitable harmonic forces the motion quickly tends to a limit cycle. The latter corresponds to the periodic motion studied by Golestanian and Ajdari \cite{9}. In our numerical work we use harmonic interactions given by the matrix
  \begin{equation}
\label{6.7}\du{K}=k\left(\begin{array}{cc}1&0\\
-1&1
\end{array}\right)
\end{equation}
with elastic constant $k$. This corresponds to nearest neighbor interactions of equal strength $k$ between the three spheres. The stiffness of the swimmer is characterized by the dimensionless number $\sigma$ defined by
  \begin{equation}
\label{6.8}\sigma=\frac{k}{\pi\eta a\omega}.
\end{equation}
We consider actuating forces oscillating at frequency $\omega$ with $(E_{1\omega},E_{2\omega})$ given by Eq. (5.11) with $\hat{\vc{\xi}}^c_\omega=\varepsilon\vc{\xi}_+$, and $E_{3\omega}=-E_{1\omega}-E_{2\omega}$. This corresponds to maximum efficiency in the bilinear theory. We choose initial conditions
 \begin{equation}
\label{6.9}x_1(0)=0,\qquad x_2(0)=d+\varepsilon a,\qquad x_3(0)=2d+\varepsilon a+\varepsilon a\mathrm{Re}\xi_+.
\end{equation}
In the bilinear theory, corresponding to small $\varepsilon$, the orbit $(r_1(t),r_2(t))=(x_2(t)-x_1(t),x_3(t)-x_2(t))$ in relative space is given by $\vc{r}(t)=\vc{r}_0+\hat{\vc{\xi}}(t)$ with $\vc{r}_0=(d,d)$ and
\begin{equation}
\label{6.10}\hat{\vc{\xi}}(t)=\varepsilon a\mathrm{Re}\vc{\xi}_+\exp(-i\omega t),
\end{equation}
independent of the elastic constant $k$.

In Fig. 1 we show the orbit for $\varepsilon=0.1$ and ratio $d/a=5$ for one period. In the figure the numerical solution cannot be distinguished from the elliptical orbit given by Eq. (6.10). By use of the Stokes parameters \cite{21} calculated from $\vc{\xi}_+$ one finds that the tilt angle in the $r_1r_2$ plane is $3\pi/4$, independent of the ratio $d/a$, and obtains an expression for the ellipticity. In the present case the ellipticity equals 0.66.

In Fig. 2 we show the orbit for $\varepsilon=2$, ratio $d/a=5$, and stiffness $\sigma=1$ for $0<t<10T$. The mean swimming velocity $\overline{U}_{sw}$ and the mean rate of dissipation $\overline{\mathcal{D}}$ are calculated as time averages over the last period $9T<t<10T$, corresponding to the limit cycle. It turns out that in the range $0<\varepsilon<2$ both quantities vary approximately in proportion to $\varepsilon^2$. In Fig. 3 we show the reduced mean swimming velocity $\overline{U}_{sw}/(\varepsilon^2\omega a)$ as a function of $\varepsilon$ for $d=5a$. In Fig. 4 we show the reduced mean rate of dissipation $\overline{\mathcal{D}}/(\varepsilon^2\eta\omega^2a^3)$ and in Fig. 5 we show the efficiency $E_T=\eta\omega a^2\overline{U}_{sw}/\overline{\mathcal{D}}$ as functions of $\varepsilon$. Interestingly, the efficiency increases monotonically with the amplitude factor.

At $\varepsilon=1.38$ we have $\overline{U}_{sw}\approx 0.023\;\omega a$ and $\overline{\mathcal{D}}\approx 47.1\;\eta\omega^2 a^3$. This can be compared with the numerical calculation of Alouges et al. \cite{22},\cite{23} on the basis of a Stokes solver. The authors used radius $a=0.05$ mm and period $T=1$ s. For viscosity of water $\eta=0.01$ poise our calculation yields $\Delta=\overline{U}_{sw}T\approx$ 0.0072 mm and $\overline{\mathcal{D}}T\approx 0.232\times 10^{-12}J$. The latter value is about the same as the one given in Table 1 of ref. 27, and the displacement agrees well with the value $0.01$ mm of Alouges et al.. A more precise comparison would require a calculation with the same hydrodynamic interactions in both procedures. In our calculation the efficiency for given amplitude $\varepsilon$ could be maximized numerically by variation of the vector $\hat{\vc{\xi}}(0)$ in the neighborhood of $\varepsilon a\mathrm{Re}\hat{\vc{\xi}}_+$ and variation of the stiffness $\sigma$. Also one can vary the equilibrium situation $\du{r}_0$ used as a starting point of the bilinear theory. The present method allows fast and straightforward design of an efficient large amplitude mechanical swimmer corresponding to an approximate form of the mobility matrix.

\section{\label{VII}Discussion}

The matrix-formulation presented above provides insight into the mathematical problem of the swimming of assemblies of spheres at low Reynolds number. It allows straightforward calculation of the swimming performance of assemblies of interest. A practical procedure for a particular swimmer would be to model the rest shape by a set of spheres, identifying the positions of the centers with the equilibrium sites of a structure with harmonic elastic interactions. By use of an approximation to the mobility matrix one can then evaluate the hermitian matrix $\du{B}$, which appears in the expression for the mean swimming velocity, and the real and symmetric matrix $\du{A}$, which appears in the expression for the mean rate of dissipation. The swimming velocity and required power can be calculated for a chosen stroke of small amplitude with harmonic time variation as expectation values of the two hermitian matrices. Large amplitudes can also be handled in principle. These require a parametric integration along a chosen closed path in the space of relative positions.

For the simple example of three collinear spheres with Oseen-type interactions, discussed in Sec. VI, the calculations can be performed in analytic form. For more complicated structures and more accurate hydrodynamic interactions the algebra rapidly becomes cumbersome, but it is straightforward to derive numerical results. Elsewhere we have applied the method to the analysis of three- and four-sphere swimmers pushing a cargo sphere \cite{24}.

As we have shown, the optimization of swimming at small amplitude leads to an eigenvalue problem. This allows determination of the optimum stroke yielding the largest swimming speed at given power. For an assembly of spheres with elastic interactions the required actuating forces which lead to optimal speed can be evaluated from the eigenvector with maximum eigenvalue. It is then of interest to study the swimming speed and power for the same set of force ratios as functions of an amplitude factor.

\newpage

\section*{Figure captions}

\subsection*{Fig. 1}
Plot of the orbit in the $r_1r_2$ plane for $d=5a,\;\varepsilon=0.1,\;\sigma=0$ for one period.

\subsection*{Fig. 2}
Plot of the orbit in the $r_1r_2$ plane for $d=5a,\;\varepsilon=2,\;\sigma=1$ for ten periods. The initial values correspond to Eq. (6.9).

\subsection*{Fig. 3}
Plot of the reduced mean swimming velocity $\overline{U}_{sw}/(\varepsilon^2\omega a)$ for $d=5a,\;\sigma=1$ as a function of the amplitude $\varepsilon$.

\subsection*{Fig. 4}
Plot of the reduced mean swimming power $\overline{\mathcal{D}}/(\varepsilon^2\eta\omega^2 a^3)$ for $d=5a,\;\sigma=1$ as a function of the amplitude $\varepsilon$.

\subsection*{Fig. 5}
Plot of the efficiency $E_T$ for $d=5a,\;\sigma=1$ as a function of the amplitude $\varepsilon$.

\newpage
\setlength{\unitlength}{1cm}
\begin{figure}
 \includegraphics{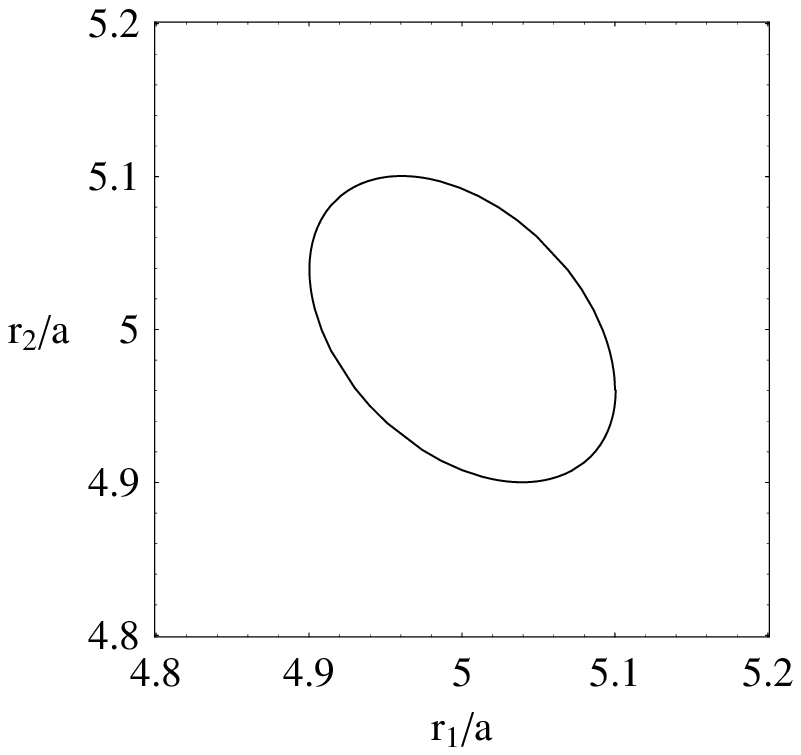}
   \put(-9.1,3.1){}
\put(-1.2,-.2){}
  \caption{}
\end{figure}
\newpage
\clearpage
\newpage
\setlength{\unitlength}{1cm}
\begin{figure}
 \includegraphics{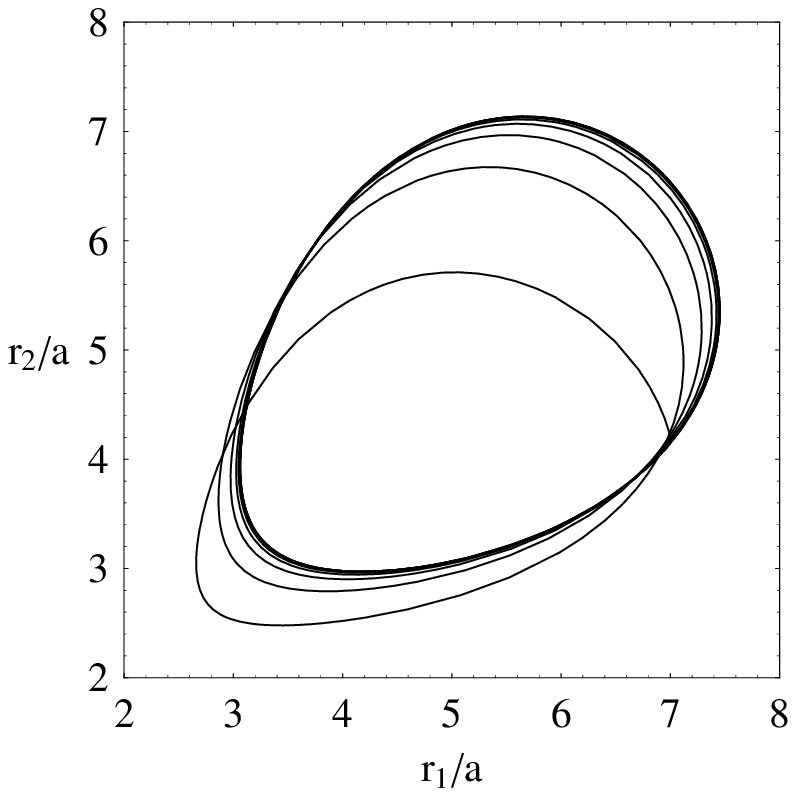}
   \put(-9.1,3.1){}
\put(-1.2,-.2){}
  \caption{}
\end{figure}
\newpage
\clearpage
\newpage
\setlength{\unitlength}{1cm}
\begin{figure}
 \includegraphics{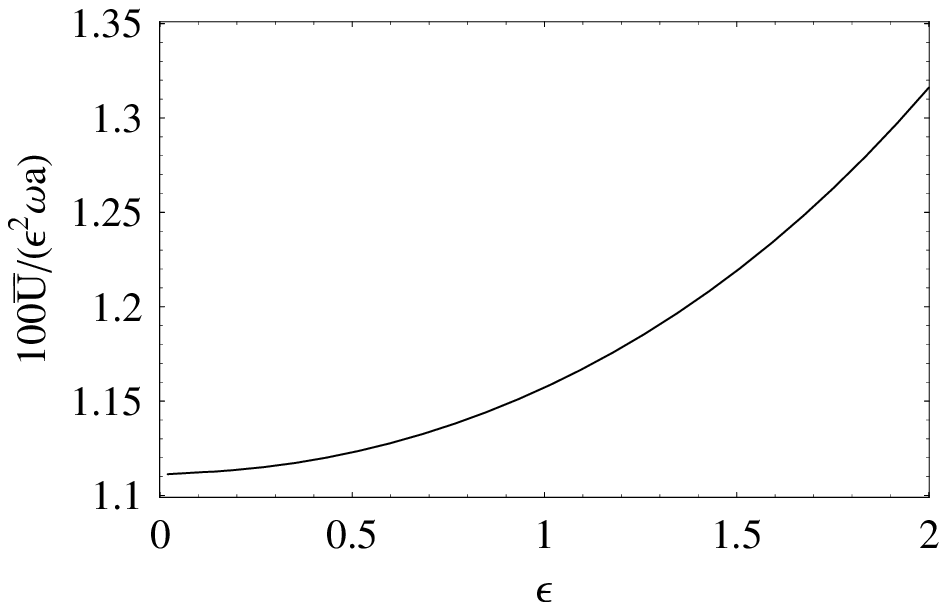}
   \put(-9.1,3.1){}
\put(-1.2,-.2){}
  \caption{}
\end{figure}
\newpage
\clearpage
\newpage
\setlength{\unitlength}{1cm}
\begin{figure}
 \includegraphics{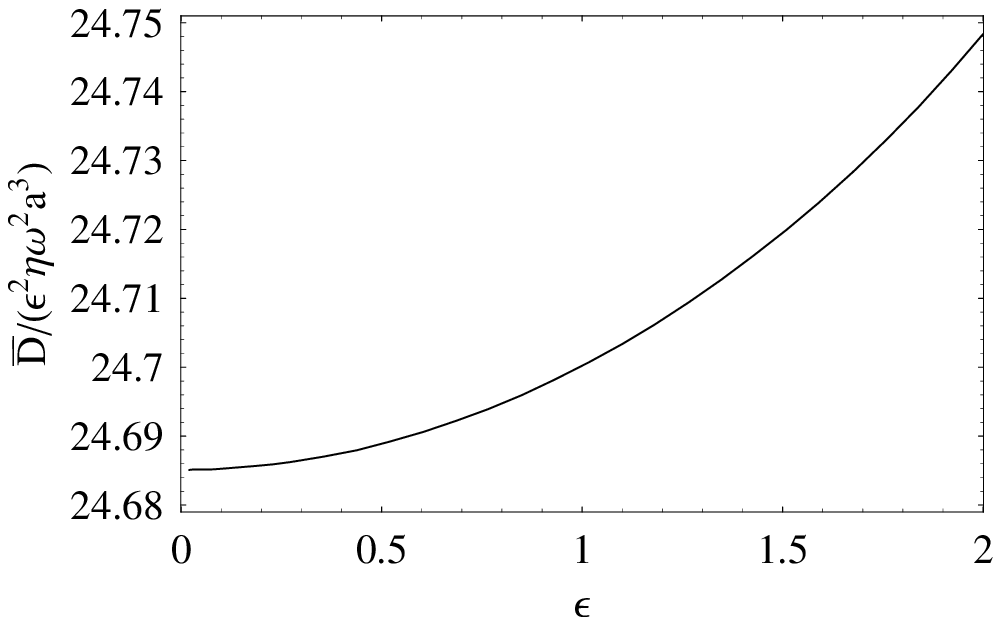}
   \put(-9.1,3.1){}
\put(-1.2,-.2){}
  \caption{}
\end{figure}
\newpage
\clearpage
\newpage
\setlength{\unitlength}{1cm}
\begin{figure}
 \includegraphics{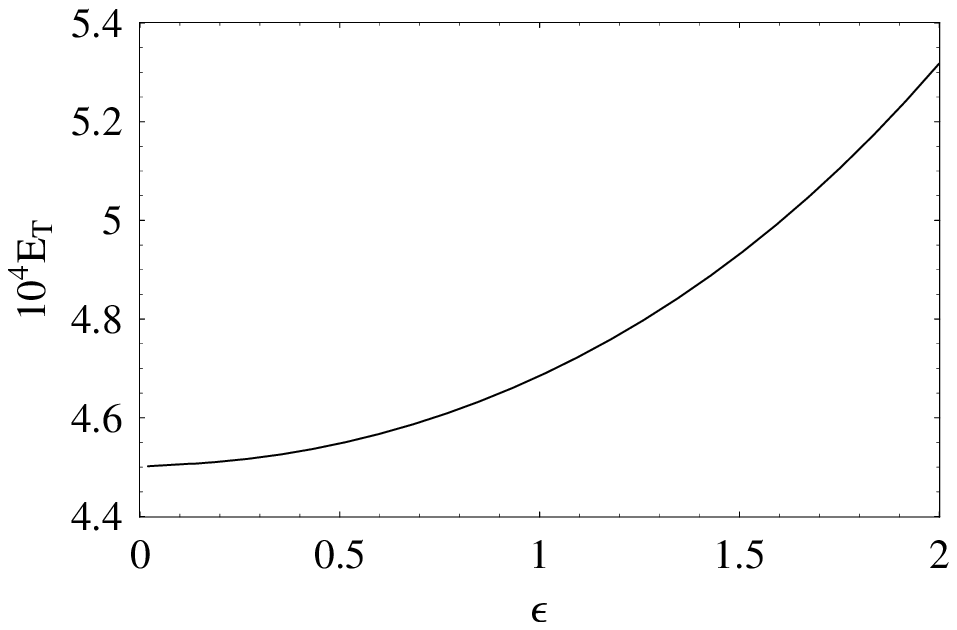}
   \put(-9.1,3.1){}
\put(-1.2,-.2){}
  \caption{}
\end{figure}
\newpage

\end{document}